\documentclass[a4paper]{article}

\usepackage{INTERSPEECH2018}
\usepackage{booktabs}
\usepackage[table,xcdraw]{xcolor}
\usepackage{dblfloatfix}

\title{A Discriminative Acoustic-Prosodic Approach for Measuring Local Entrainment}
\name{Megan M. Willi$^1$, Stephanie A. Borrie$^2$, Tyson S. Barrett$^2$, Ming Tu$^1$, Visar Berisha$^1$}
\address{
  $^1$Arizona State University, Tempe, AZ, USA\\
   $^2$ Utah State University, Logan, UT, USA}
\email{megan.willi@asu.edu, \{stephanie.borrie, tyson.barrett\}@usu.edu, \{mingtu, visar\}@asu.edu}

\begin{document}

\maketitle
\begin{abstract}
Acoustic-prosodic entrainment describes the tendency of humans to align or adapt their speech acoustics to each other in conversation. This alignment of spoken behavior has important implications for conversational success. However, modeling the subtle nature of entrainment in spoken dialogue continues to pose a challenge. In this paper, we propose a straightforward definition for local entrainment in the speech domain and operationalize an algorithm based on this: acoustic-prosodic features that capture entrainment should be maximally different between real conversations involving two partners and sham conversations generated by randomly mixing the speaking turns from the original two conversational partners. We propose an approach for measuring local entrainment that quantifies alignment of behavior on a turn-by-turn basis, projecting the differences between interlocutors' acoustic-prosodic features for a given turn onto a discriminative feature subspace that maximizes the difference between real and sham conversations. We evaluate the method using the derived features to drive a classifier aiming to predict an objective measure of conversational success (i.e., low versus high), on a corpus of task-oriented conversations. The proposed entrainment approach achieves 72\% classification accuracy using a Naive Bayes classifier, outperforming three previously established approaches evaluated on the same conversational corpus. 
   
\end{abstract}

\section{Introduction}
\vspace{-0.2cm}
Conversational entrainment describes a communication phenomenon in which humans align their behaviors to each other during conversation. This alignment has been shown to be important for successful conversation. In summarizing some of the key literature investigating entrainment and conversational success, Borrie and Liss concluded that entrainment serves as a powerful coordinating device that impacts both comprehension and the development of positive social relationships \cite{borrie_rhythm_2014}. Acoustic-prosodic entrainment, an aspect of conversational entrainment, refers to the alignment of speech features. For example, people align the pitch and intensity of their speech to closely match those features in the speech of their conversational partner \cite{borrie_disordered_2015}.

A body of literature investigating acoustic-prosodic entrainment exists; however, a consensus on how entrainment is measured or quantified at the local turn-by-turn level does not. Local entrainment refers to entrainment measures at the speaking turn level while global entrainment refers to entrainment measures at the speaker level \cite{lee_computing_2014}. Broadly speaking, the literature reports on two main approaches to measuring local acoustic-prosodic entrainment. One line of research \cite{levitan_measuring_2011} uses domain knowledge to develop interpretable, acoustic-prosodic entrainment features in real conversations. Generally these approaches use simple correlation measures such as synchrony, or the Pearson correlation between separate speakers' acoustic features across a speaking turn, as acoustic features. These measures are then validated by demonstrating differences between these same measures derived from real and sham conversations. The second line of research  \cite{nasir_predicting_2017} is more data-driven and develops complex features that capture the relationship between conversational partners' acoustic-prosodic features, but for the purpose of predicting a specific outcome measure, such as marriage success in a Couple Therapy corpus. Often these approaches use projection measures rather than simple correlation measures. For example, \cite{lee_computing_2014} used a principal components analysis (PCA) to project the original acoustic-prosodic features into a new, projected subspace.

Our approach uses a unique combination of the previous methods to develop a new measure of entrainment using a two-step process: step one exploits the differences between real and sham conversations to develop acoustic-prosodic features of entrainment and step two validates these features by seeing how well they predict an aspect of conversational success. We use linear discriminant analysis (LDA) to quantify acoustic-prosodic entrainment on a turn-by-turn basis, projecting the differences between interlocutors' acoustic-prosodic features for a given turn onto a discriminative feature subspace that maximizes the differences between real and sham conversations. The resulting projected acoustic-prosodic feature is motivated by the assumption that acoustic-prosodic features that capture turn-by-turn entrainment should be maximally different between real and sham conversations between the same two individuals. Using real and sham conversations to develop (rather than validate) new acoustic-prosodic features of entrainment is a novel approach and a key scientific contribution of this paper. A small set of entrainment features generated in this way from four larger feature sets are used to predict an objective measure of conversational success; communicative efficiency. This objective measure of conversational success has been previously correlated with acoustic-prosodic entrainment \cite{borrie_disordered_2015}\cite{s._a._borrie_conversational_2017}. Our proposed approach is compared with three additional approaches that have been used to measure local entrainment measures in acoustic-prosodic behavior.       

\begin{figure*}[!b]
  \centering
  \includegraphics[scale=.30]{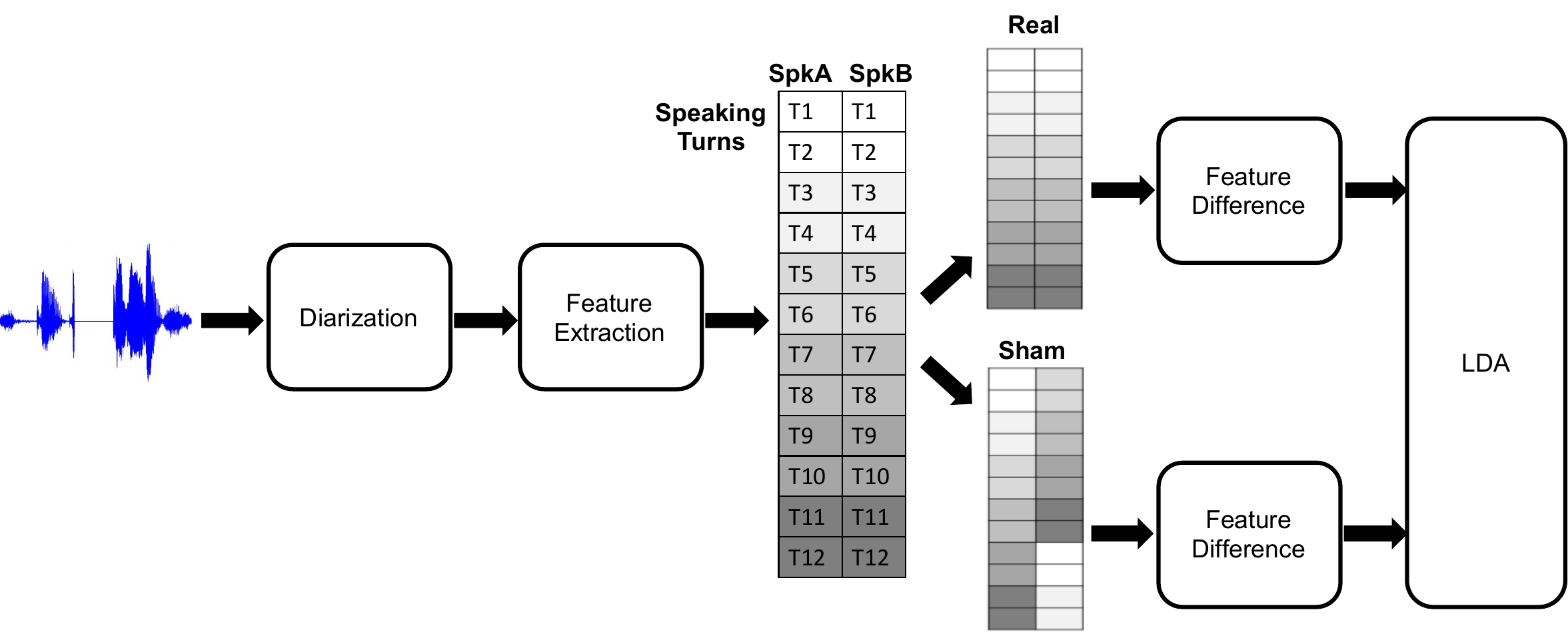}
   \caption{An overview of the proposed local entrainment analysis framework. The main steps include speaker diarization, feature extraction, creation of the sham conversations, extraction of the speaker difference measures, and LDA analysis. The grey scale indicates the temporal order of the speaking turns (T).}
  \label{fig:LDA_analysis}
\end{figure*}
\section{Proposed Approach}

\subsection{Measuring Local Entrainment using LDA}
In the current study, we use LDA \cite{fisher_use_1936} to maximize the separation between the original conversations and their sham counterparts. This allows us to project the data into a new feature space that maximally differentiated between real and sham conversations. The purpose of the projection was to 1) ensure that all conversations could be compared in the same feature space and 2) operationally define local entrainment as the projected feature representation that maximally differentiates between real and sham conversations. An overview of the processing steps are graphically depicted in Figure~\ref{fig:LDA_analysis}-- we start with an audio recorded conversation, diarize it and extract acoustic-prosodic features at the turn level; then we form a sham version of the original conversation and measure the feature difference between consecutive turns of conversational partners. Finally, these features are used to learn a discriminative feature subspace that maximally separates between real and sham conversations. Details regarding the specifics of this approach are outlined below.  
                                          
\subsubsection{Sham Conversations}

Our approach requires that the conversation is diarized at the level of a speaking turn (either automatically or manually). Sham conversations are randomized versions of the original conversations that disrupt the natural, turn-by-turn dialogue exchange in conversations. For example, \cite{lee_computing_2014} used randomly generated dialogues between partners from different conversations to validate an approach for measuring entrainment. The current study uses a more specific type of sham conversation, similar to \cite{arthur_ward_measuring_2007}, which also disrupts the turn-by-turn modulation of behavior but retains acoustic-prosodic data from both original partners. Rather than completely randomizing each conversational partner's turns, each partner's turns were first divided into thirds and then the order between turn blocks was block-randomized (see sham in Figure~\ref{fig:LDA_analysis}). This randomization scheme was employed to maximally preserve the within-speaker dyadic coordination while disrupting the block order of at least one partner's turns. Each speaker block was paired with each non-corresponding speaker block (i.e., AB and AC, but not AA) resulting in two within-conversation shams per dyad (i.e., a total of 106 shams).

\subsubsection{Feature Extraction and LDA Analysis}
\begin{figure}[t]
  \centering
  \includegraphics[scale=.20]{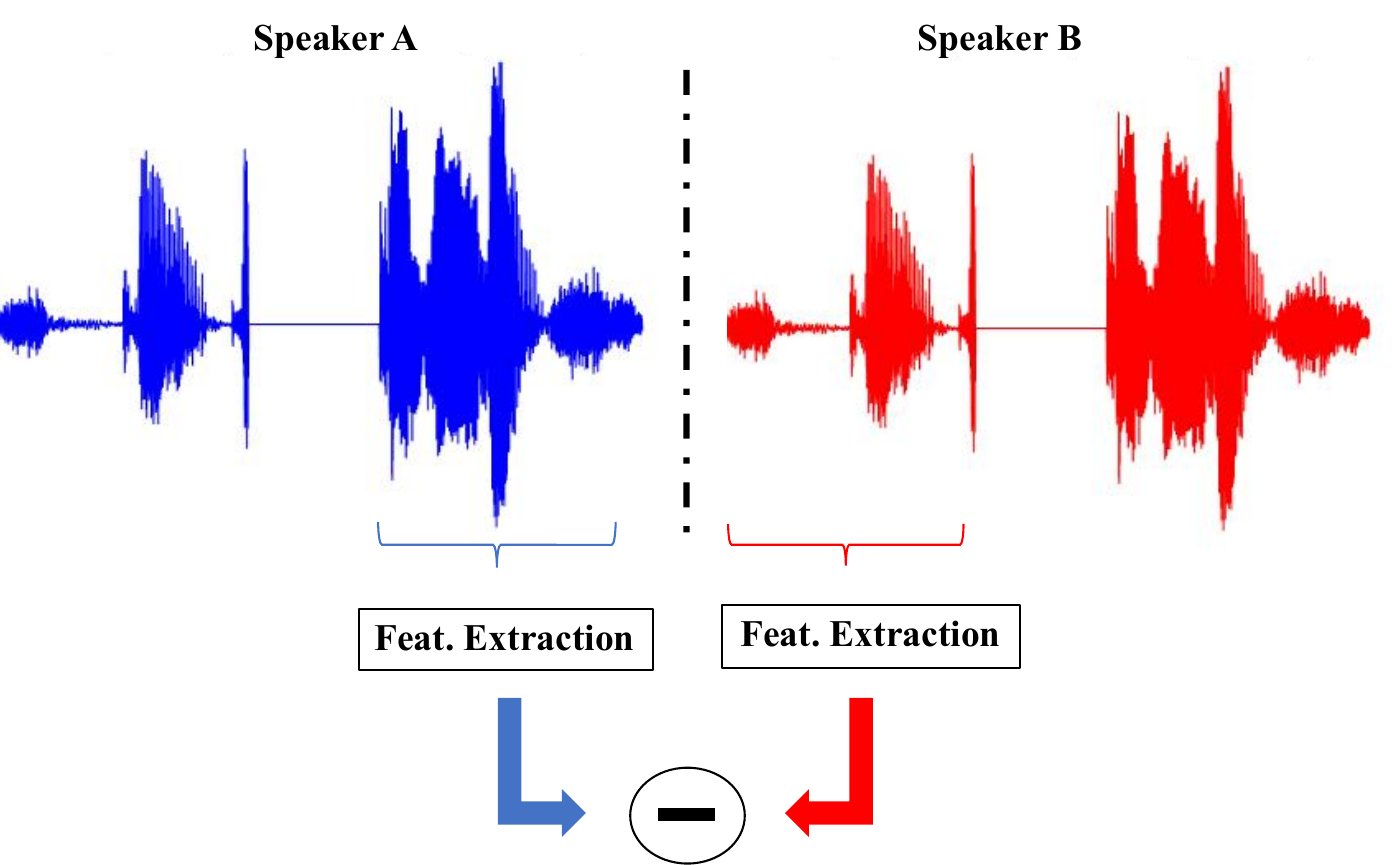}
  \caption{Features are extracted from the speaker utterances adjacent a change in speaking turn. The difference measure for a given turn is the absolute difference between these features.}
  \label{fig:diff_measure}
\end{figure}

A set of features are extracted for each speaking turn. In our implementation we extract four different feature sets (i.e., MFCC Statistics, EMS, LTAS, and Phonation Features; see Section 3.3 for details).  For each conversation, the magnitude of the feature difference between the utterances immediately adjacent to a change in speaking turn (i.e., as designated by the utterances' turn-midpoints) was calculated. The result was a difference measure for each feature for each speaking turn in a given conversation. This idea is graphically depicted in Figure~\ref{fig:diff_measure}. Next, the difference measures within each conversation were grouped into their corresponding feature sets (Section 3.3). The result was four groups, designated by feature set, that included all the feature difference measures for all the speaking turns. This process was repeated for both the real and sham conversations, but the designation between the real and sham groups was preserved.

The LDA analysis was performed on each feature set separately. For the analysis, the difference measures were estimated for every change in speaking turn and divided into two classes: real and sham. Let $x_{i}$ denote a feature difference matrix for speaking turn change $i$ and let $y_{i}$ represents the class (real vs. sham) for the same sample. We define the within-class scatter matrix as,
 
\begin{equation}
  S_{w}=\displaystyle\sum_{i=1}^{n} (x_{i}-\mu_{y_{i}})(x_{i}-\mu_{y_{i}})^{T},
  \label{eq1}
\end{equation}
Where $n$ is the number of speaking turns. Here, $\mu_{y_{i}}$ is the sample mean of the class (real or sham) represented by $y_i$. The between-class scatter matrix is defined as:

 \begin{equation}
  S_{b}=\displaystyle\sum_{k=1}^{2} n_{k}(\mu_{k}-\mu)(\mu_{k}-\mu)^{T}.
  \label{eq2}
\end{equation}

Here, $\mu$ is the overall sample mean, and $n_{k}$ is the number of samples in the $k$-th class. Then, LDA can be formulated as an optimization problem to find a projection,  $w$, that maximizes the ratio of the between-class scattering to the within-class scattering, as

 \begin{equation}
  \hat{\mathrm{w}}=\underset{\mathrm{w}} {\arg \! \max} \frac{\mathrm{w}^{T}S_{b}\mathrm{w}}{\mathrm{w}^{T}S_{w}\mathrm{w}}.
  \label{eq3}
\end{equation}

The solution is given by the following generalized eigenvalue problem:
 \begin{equation}
  S_{b}\mathrm{w}=\lambda S_{w}\mathrm{w}.
   \label{eq4}
\end{equation}

The result of this process is a projection matrix that transforms the high-dimensional features into a single feature that maximizes the difference between real and sham conversations. This process is performed independently for each of the four feature sets considered in this paper. This means that, for every feature set, a single LDA projection matrix is derived and difference features at the turn level can be projected on this matrix. This results in four LDA features being extracted for each conversation at the level of a turn. We convert the turn-level LDA features to conversation-level features by calculating the min, max, mean, and standard deviation across the conversation. This results in a 16-dimensional feature vector for each conversation (4 statistical features for each of the 4 LDA feature sets).
  
\section{Data and Features}

\subsection{Dataset}
This work is based on a corpus of 53 task-oriented conversations, elicited from students engaged in university-level education. Conversations consisted of two partners paired up to form a dyad, and thus the corpus involved 106 conversational partners (93 females and 13 males) aged 19--28 years old. All conversational partners were native speakers of American English with no history of speech, language, hearing, or cognitive impairment. Each dyad participated in the Diapix Task, a cooperative task where partners verbally communicate to identify differences (as quickly and as accurately as possible) between two different versions of the same picture scene \cite{van_engen_wildcat_2010}. Details of the interaction task, instructions and recording equipment are specified in \cite{borrie_disordered_2015}. Each conversation consisted of 11-14 minutes (M = 11.5 mins) of spoken dialogue. 

An objective measure of an aspect of conversational success was used as the prediction target for the classification task. This measure was based on the total the number of differences identified during the first 10 minutes of the conversation task. Thus, the measure represents a simple, gross measure of communicative efficiency: relatively low and high numbers of identified differences indicate relatively high and low communicative efficiency, respectively. This prediction target measure has been found to track with entrainment index scores using previous approaches of quantifying entrainment \cite{borrie_disordered_2015}\cite{s._a._borrie_conversational_2017}; and is presumably influenced by fewer extraneous factors than domain-specific measures such as marital success.

We use the entrainment features to predict a binary measure of conversational success. The total number of differences identified by dyads in the task-oriented conversations ranged from 10 to 30. Using a binary class label, the conversations were classified as having low efficiency (i.e., 10-19 differences) or high efficiency (i.e., 21-30 differences). From the corpus of 53 conversations, 27 conversations were classified as low and 23 conversations were classified as high. The three conversations that scored 20 differences (median score) were excluded from the classification analysis to avoid class overlap and the use of a continuous scale.   

\subsection{Preprocessing of Audio Data}
The following section describes the pre-processing steps used to prepare the recorded conversations for the feature extraction and subsequent entrainment analyses. Trained research assistants manually annotated the conversations, by speaker, into spoken utterances using Praat \cite{boersma_praat:_2014} acoustic analysis software. A spoken utterance is defined as an inter-pausal unit (IPU) from a single speaker, including pauses of up to 0.5 seconds \cite{levitan_measuring_2011}. Any non-speech section less than 0.5 seconds is included in the spoken utterance that surrounds it and spoken utterances that overlap in time are permitted. 

All sound files were normalized using a standard loudness normalization procedure based on a reference level, ensuring that the resulting normalized signal was within -1 to 1 to avoid peak clipping. Additionally, all sound files were downsampled to 16kHz prior to feature extraction.

\subsection{Feature Extraction}
The following acoustic-prosodic features were extracted for each spoken utterance. This resulted in a 418-dimensional feature vector for each utterance. Similar feature extraction methods have been used previously in \cite{berisha_modeling_2014}\cite{tu_models_2016}\cite{tu_objective_2017}\cite{ming_tu_interpretable_2017}. Additional post-processing prior to entrainment feature extraction included removing any utterance that was 0.5 seconds or less and by replacing any missing values with the median feature value for that conversation.

\subsubsection{MFCC Statistics}
The Mel-Frequency Cepstral Coefficient (MFCC) statistics are calculated from the 13th order MFCCs (including 0th order) and their first and second order derivatives. The MFCCs are extracted using a 20 ms window with 10 ms frame increment. Then, across all frames of MFCCs, the mean, standard deviation, range, skewness, kurtosis and mean absolute deviation are calculated for each dimension. This results in a 234 (6x39)-dimensional feature vector.

\subsubsection{EMS}
The Envelope Modulation Spectrum (EMS) measures long-term changes in energy distribution across various frequency bands \cite{liss_discriminating_2010}. The original speech segment, $x_{0}(n)$, is first filtered into 9 octave bands with center frequencies of approximately 30, 60, 120, 240, 480, 960, 1920, 3840, and 7680 Hz using eight-order Butterworth filters. Let $h_{i}[n]$ denote the filter associated with the $i$th octave. The filtered signal $x_i[n]$ is then denoted by,

\begin{equation}
  x_{i}[n]=h_{i}[n]*x[n], i=1,...,9.
  \label{eq1}
\end{equation}

The envelope for the ten signals (the original signal and 9 octave band signals), denoted by $e_{i}[n]$, is extracted by:

\begin{equation}
  e_{i}[n]=h_{LPF}[n]*|\mathcal{A}\{x_{i}[n]\}|, i=1,...,9.
  \label{eq2}
\end{equation}

where, $\mathcal{A}\{x(t)\}=x[n]+j\mathcal{H}\{x[n]\}$ is the analytical signal, $h_{LPF}[n]$ is the impulse response of a fourth-order, 30 Hz low-pass Butterworth filter, and $\mathcal(H)\{\cdot\}$ denotes the Hilbert transform.

Once the amplitude envelope of each signal is obtained, the mean is removed and the power spectrum for each of the bands, $PowSpec_{i}$, is estimated by evaluating the DFT using the Goertzel algorithm at frequencies $0 Hz < f \leq 10 Hz$. From the power spectrum, six EMS metrics are computed for each of the 9 octave bands, and the full signal (see \cite{noauthor_notitle_nodate} for more details). This results in a 60-dimensional feature vector.

\subsubsection{LTAS}
The Long-Term Average Spectrum (LTAS) \cite{mendoza_differences_1996} captures the average spectral information in each octave of the signal. 

The original speech segment, $x_0[n]$, is first filtered into 9 octave bands using the same filter bank settings as in the EMS feature extraction. Each of the ten band signals (the original full-band signal and 9 octave band signals), $x_i [n],\ i = 0,\ldots,9$, is then framed using a 20 ms non-overlapping rectangular window and the Root Mean Square (RMS) of each frame is estimated. Finally, ten features are extracted for each of the $i$ bands (see \cite{noauthor_notitle_nodate} for more details). This results in a 99-dimensional feature vector.

\subsubsection{Phonation Features}
Praat \cite{boersma_praat:_2014} was used to automatically extract phonation features including fundamental frequency, jitter, shimmer, and Harmonics-to-Noise Ratio (HNR). The features respectively provide information about pitch, cycle-to-cycle pitch variation, cycle-to-cycle amplitude variation, and an estimate of the noise level in the human voice. Results from the phonation measurements depend on the pitch extraction settings. The phonation features were extracted using a 5 ms time step and default parameters for pitch floor, pitch ceiling, silence threshold, and voicing threshold. Additional voice quality features as well as measures of central tendency and variation were also included in the feature set resulting in a 24-dimensional feature vector.

\subsubsection{Intensity}
OpenSmile \cite{florian_eyben_opensmile_2010} was used to automatically extract the mean intensity resulting in one additional feature. The intensity feature was included for comparison purposes with previous local acoustic-prosodic entrainment measures. The parameters used to extract the intensity feature are described in \cite{nichola_lubold_acoustic-prosodic_2014}.   

\section{Results}
We compare the proposed approach against three existing methods for measuring local acoustic-prosodic entrainment. Three machine learning classifiers were used to compare performance between the four local measures: Logistic Regression, Support Vector Machine, and Naive Bayes. The classifiers were chosen because they have been used most recently in the literature for prediction tasks related to acoustic-prosodic entrainment \cite{nasir_predicting_2017} and because they provide a descriptive comparison between linear and non-linear classification methods. The classification experiments were conducted using leave-one-out cross validation. All classifiers were implemented in WEKA with default settings \cite{mark_hall_weka_2009}. The three existing methods of measuring local entrainment include:

\textbf{\textit{Convergence, Proximity, Synchrony:}} Local synchrony, convergence, and proximity measures were all calculated as described in \cite{levitan_measuring_2011}. Herein, we compute these measures for mean and max pitch, local jitter, local shimmer, mean intensity, and mean HNR descending. This results in an 18-dimensional feature vector for each conversation.

\textbf{\textit{PCA Symmetric Similarity:}} Local PCA similarity measures were computed as described by \cite{lee_computing_2014} regarding their global PCA similarity measures, except, instead of a single score per conversation, we split the conversation into the first and second halves. We then obtained a PCA similarity score for each half. A PCA similarity score was extracted for EMS, MFCC, LTAS, and Phonation Features for each conversation, resulting in a four-dimensional feature vector for each conversation.

\textbf{\textit{Short Term Dynamic Functionals:}} Short term dynamic functionals capture the dynamic interaction between conversational partners in conversation. The differences (i.e., deltas) between corresponding features in adjacent speaker turns were calculated following the procedures outlined in \cite{nasir_predicting_2017}. The turn-level analysis described in \cite{nasir_predicting_2017} was adapted to an utterance level analysis and statistical functionals (i.e., mean, median, and standard deviation) of the utterance-level delta features were selected using a simple correlation-based feature selection method, where the features with the highest Pearson's correlation coefficient to the outcome measure were selected. In the present study, a 15-dimensional feature vector for each conversation was extracted, approximating the feature vector length (i.e., 10-20\% of the 74 acoustic features) outlined in \cite{nasir_predicting_2017}.    

\begin{table}[]
\centering
\caption{Entrainment Classification Performance}
\label{tab:results}
\begin{tabular}{@{}llll@{}}
\toprule
\textit{\textbf{Entrainment}} & \textbf{Logistic} & \textbf{SVM}   & \textbf{Naive Bayes} \\ \midrule
\textit{LDA}                  & 56.00             & \textbf{68.00} & \textbf{72.00}       \\
\textit{Prox/Conv/Sync}       & \textbf{64.00}    & 51.20          & 50.00                \\
\textit{PCA}                   & 56.00             & 31.00          & 42.00                \\
\textit{STDF}                 & 44.00             & 61.60          & 52.00                \\  \bottomrule
\end{tabular}
\end{table}

\subsection{Classification Performance}

The results of the classification experiments are shown in Table~\ref{tab:results}. The LDA entrainment approach using a Naive Bayes classifier revealed the highest classification performance at 72\%. In addition, the LDA entrainment measures outperformed the other entrainment approaches in two out of the three classifiers. The proximity, convergence, and synchrony measures outperformed the alternative entrainment approaches when using a linear logistic regression classifier, but the LDA entrainment approach outperformed these results by 4\% and 8\% when using Support Vector Machine and Naive Bayes classifiers, respectively.    

\vspace{-0.2cm}
\section{Conclusion}

We present a new method for capturing local entrainment in acoustic-prosodic behavior that is motivated by the straightforward observation that entrainment values should be higher in real versus sham conversations and should predict a presumed functional outcome of entrainment, conversational success. Our unique discriminative approach of using real and sham conversations to create turn-level, entrainment features resulted in improved classification performance. While applied to acoustic-prosodic speech features in the current paper, our approach provides a general framework for measuring local entrainment that can generalize to other modalities. For example, local lexical entrainment could be quantified as the root mean squared error between real and sham linguistic influence matrixes \cite{ kawabata_lexical_2016}. Thus, future studies could examine this discriminative approach for quantifying entrainment with other aspects of communication. 

\vspace{-0.2cm}
\section{Acknowledgements}
This research was supported by the National Institute of Deafness and Other Communication Disorders, National Institutes of Health Grants R21DC016084-01 and R01DC006859. 

\bibliographystyle{IEEEtran}

\bibliography{Interspeech_2018_Entrainment_Final}

\end{document}